# First principles study of oxidation of Si segregated $\alpha$-Ti(0001) surfaces

**Somesh Kr. Bhattacharya**[*], Ryoji Sahara, Tomonori Kitashima

Research Center for Structural Materials, National Institute for Materials Science, 1-2-1 Sengen, Tsukuba, Ibaraki 305-0047, Japan

Kyosuke Ueda, Takayuki Narushima

Department of Materials Processing, Tohoku University, 6-6-2 Aza Aoba, Aramaki, Aoba-ku, Sendai 980-8579, Japan

**Abstract**

The oxidation of α-Ti(0001) surface was studied using density functional theory. In order to enhance the oxidation resistance, we substituted Ti atoms with Si atoms in Ti(0001) surface. We observe that Si prefers to segregate at the surface layer of Ti(0001) compared to other sub-surface layers. The Si solubility on the Ti(0001) surface was estimated to be around 75 at.%. While Si segregation was found to reduce the binding between the oxygen atom and the Ti(0001) surface, the barrier for oxygen to diffuse into the slab increases. The dissociation of the oxygen molecule on the clean and Si segregated surfaces of Ti was found to be barrier-less. Overall, the Si segregation on the Ti(0001) surface was found to hinder the dissolution of oxygen in Ti.

## (1) Introduction

Titanium (Ti) exhibits superior mechanical properties while being extremely light weight. Consequently, Ti and Ti alloys are widely used in the aviation and aerospace sector, in high performance sports equipment, and in the medical field for bone implants and replacement devices [1]. The ability to withstand moderately high temperature without creeping makes Ti alloys a suitable material for components in jet engines. However, at higher temperature, the deterioration due to oxidation of these alloys affects the critical components of jet engines, especially those with thin sections. The oxidation can lead to surface embrittlement and degrade the mechanical properties of the alloys [2]. Surface modification processes to improve the oxidation resistance of Ti alloys have been studied. These include surface plating, plasma spraying, chemical or physical vapor deposition or ion implantation. Most of these processes not only require special and expensive equipments but also have relatively high treatment cost and low productivity [3-7]. Thus, the oxidation of Ti alloys and an effective way to counter it remains a challenging problem till date.

Oxidation of Ti and Ti alloys have been extensively studied over the past several decades. The oxidation of Ti in the temperature range of 800-1200ºC was studied by Stringer [8] and Kofstad et al [9]. Oxidation of clean Ti(0001) surface at room temperature was studied by Shih et al. using low energy electron diffraction (LEED) and Auger electron spectroscopy (AES) [10]. From the AES spectra, they concluded that the oxidized state is probably TiO and not $TiO_2$. The oxidation kinetics of Ti films was studied by Cichy et al. at 300 K using quartz crystal microbalance [11]. Lu et al. studied the oxidation of Ti surface by oxygen and water [12]. They observed that even at 150 K oxygen oxidizes $Ti^0$ to $Ti^{4+}$, $Ti^{3+}$ and $Ti^{2+}$ state. David et al. studied the oxygen diffusion in α-Ti using nuclear microanalysis [13]. Kitashima et al. studied the effect of alloying elements on the tensile and oxidation properties of α and near-α Ti alloys. They observed that while Ge and Sn decrease the oxidation resistance in α-Ti, the addition of Zr, Hf, Si or Nb is beneficial towards oxidation resistance [14]. The role of these elements towards the oxidation resistance of Ti was outlined by Kitashima and Kawamura [15]. Especially, Si is known to effectively lower the oxidation rate of Ti [16-18]. Chaze et al. pointed out that Si prevents the formation of oxide ($TiO_2$) on Ti and the dissolution of oxygen in Ti [16], which are the main processes in the oxidation of Ti at elevated temperatures. Dai et al. reviewed the high temperature oxidation of Ti alloys and Ti aluminides [19].

Atomistic simulations using classical force fields and density functional theory (DFT) have also been performed to understand the oxidation behavior of Ti. Schneider et al. combined first principles molecular dynamics (FPMD) and classical force field to study the Ti/TiO$_x$ interface [20]. On the other hand Ohler et al. used DFT to study the TiO$_2$/Ti interfaces [21]. Their calculated work of separation shows strong binding between the metallic and the oxide layer. Wu et al. studied in detail the diffusion of oxygen in α-Ti [22] and the effect of substitutional impurities on the diffusion of oxygen in α-Ti [23]. The oxygen diffusion in the Ti(0001) was studied by Liu et al. using DFT [24]. The authors calculated the barrier for oxygen diffusion from surface to the sublayers of Ti(0001) using the nudged elastic band method (NEB).

Despite the available literature, our knowledge of Ti oxidation remains limited. Although experimentally the effect of alloying elements has been studied, very little has been done from the theoretical point of view. Hence, in this work we performed DFT simulations to understand the effect of alloying on the dissolution of oxygen on Ti surfaces. We studied in detail the oxygen adsorption, oxygen diffusion and dissociation of oxygen molecule on Ti(0001) surfaces alloyed with Si.

**(2) Methodology**

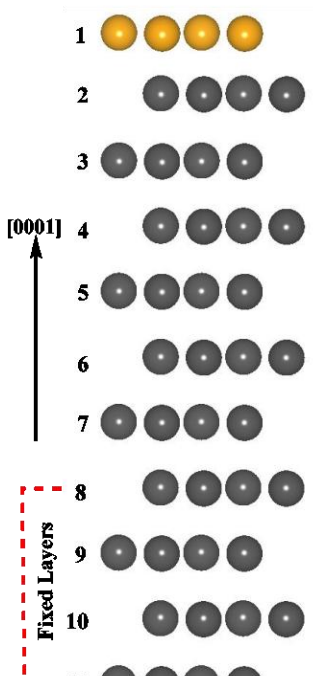

**Figure 1** The α-Ti(0001) slab model. The surface layer is labeled as '1' while the bottom most layer is labeled as '11'. The fixed layers are shown by the broken line.

We performed the DFT calculations using the Vienna *Ab initio* Simulation package (VASP) [25, 26]. We used the projector augmented wave (PAW) method [27, 28] to represent the interactions between electrons and ions while the exchange-correlation was described using the Perdew-Burke-Ernzerhof (PBE) functional [29]. The electronic wavefunction was expanded using a plane wave basis with an energy cut-off of 500 eV while the Brillouin zone (BZ) was sampled using the Monkhorst-Pack scheme [30]. The geometry optimization was performed using the conjugate gradient (CG) scheme where the convergence threshold for energy and atomic forces was set to $10^{-6}$ eV and $10^{-2}$ eV Å$^{-1}$, respectively. Our calculated lattice parameters $\vec{a}$ and $\vec{c}$ for bulk α-Ti are 2.922 Å and 4.631 Å, respectively. These are in good agreement with previously reported values [24, 31-33].

We constructed an asymmetric eleven layer Ti slab to represent the α-Ti(0001) surface as shown in Fig. 1. The four bottom layers of the slab were kept fixed at the bulk distance while other atoms were allowed to relax. A vacuum of 20 Å was introduced along the [0001] direction to avoid any spurious interaction. The Si segregation was studied by substituting Ti with Si atom. Si atoms were substituted at different layers to understand the preferred substitution sites. To generate systems with different Si concentration, we not only varied the number of Si atoms in the cell but also the slab size from (1x1) to (4x4). To study the oxidation process, we adsorb oxygen atoms and oxygen molecule on the clean and Si segregated Ti slabs. We used the NEB method [34] to calculate the oxygen diffusion barrier. The energy barriers were calculated by generating the potential reaction path for the oxygen diffusion with a series of intermediate images. The intermediate images were relaxed such that the *z*-coordinate (*z*-axis is parallel to the [0001] axis) of the oxygen atom was kept fixed while all other coordinates were allowed to relax.

## (3) Results

### (3.1) Si segregation in Ti(0001)

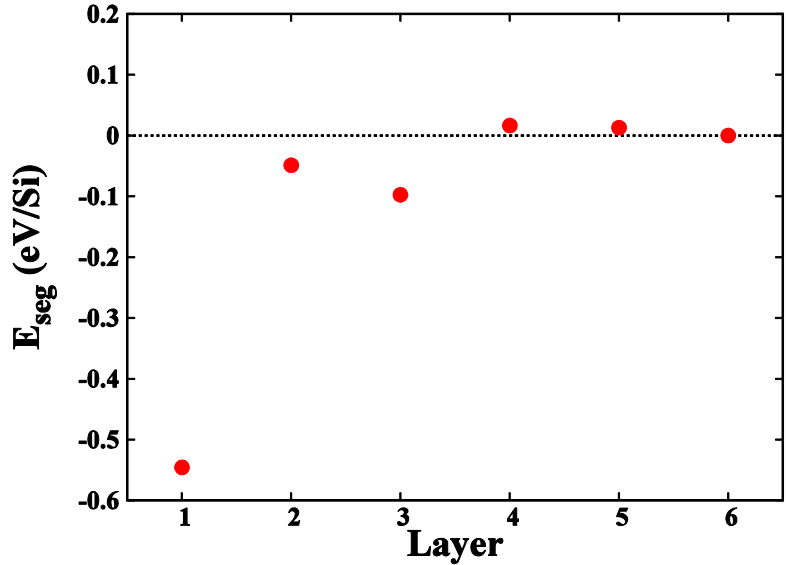

**Figure 2** Variation of the Si segregation energy ($E_{seg}$) among the different layers of the slab.

To estimate the most favorable site of Si segregation in the Ti slab, we considered the (2x2) Ti slab and substituted one Ti with Si at various layers. The segregation energy per Si atom ($E_{seg}$) is defined as the difference in the total energy between the systems with Si substituted in a particular layer and at bulk-like site (in our case in the middle of the slab; layer '6') [35, 36]. A negative $E_{seg}$ indicates favorable segregation and vice versa. The variation of $E_{seg}$ for Si in different layers of the Ti slab is shown in Fig. 2. For layer 1, corresponding to surface segregation, the $E_{seg}$ is -0.55 eV/Si while for layers 2 and 3 the $E_{seg}$ is -0.05 eV and -0.1 eV, respectively. For other deeper sites (in layers 4 and 5), the segregation is not favorable. Thus, for surface segregation the energy gain is large compared to subsurface or deeper layers. Therefore, in this work, we only deal with the Si segregated cases where Si occupies the surface positions on the Ti(0001) slab model.

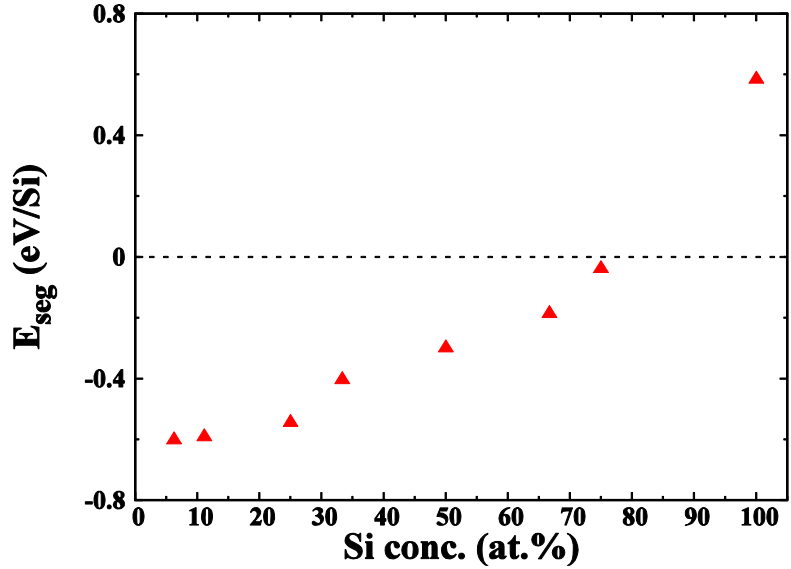

**Figure 3** Variation of the segregation energy as a function of Si concentration at the surface layer of α-Ti(0001).

It is well known that the solubility of Si in bulk α-Ti is very low [15]. In order to estimate the solubility of Si on the (0001) surface of α-Ti, we considered various supercells of the slab model as mentioned in section 2. At the surface layer of these supercells, we substituted Ti atoms by Si atoms to generate various Si concentration cases. The Si concentration is defined as the ratio of the number of Si atoms to the total number of atoms in the surface layer. In this case, the segregation energy $E_{seg}$ is calculated as difference in the total energy between systems with Si in a surface site and a bulk-like site (in our case in the middle of the slab; layer '6') [35, 36]. The variation of $E_{seg}$ for different Si concentration is shown in Fig. 3. It has already been mentioned earlier that negative $E_{seg}$ indicates stable Si segregation. As illustrated in Fig. 3, when the Si concentration is decreased from 100 at.%, the magnitude of $E_{seg}$ decreases monotonically and becomes negative at for all Si concentrations ≤ 75 at.% indicating favorable segregation. Thus, the solubility of Si on α-Ti(0001) surface is around 75 at.%. This is an important result of this work as no such experimental or theoretical data is available till date. For all further calculations in this work, we have considered Si concentrations less than 75 at.%.

The interaction between the Si and the Ti atoms on the surface can be understood from the density of states (DOS). Figure 4 shows the projected DOS (PDOS) plot of the Si-*s* and *p* and the *d* states of surface Ti atoms for the Si segregated Ti surface with 25 at.% Si. The PDOS plot shows overlap between the Si-*p* states and Ti-*d* states is significant in the energy range of -4 eV to the Fermi level ($E_F$) (set at zero). This indicates a *p-d* hybridization which stabilizes the Si segregation on the surface of Ti(0001). However, it is important to note that the Si-*p* states spread over an energy range of approximately 4 eV and have very small peak heights.

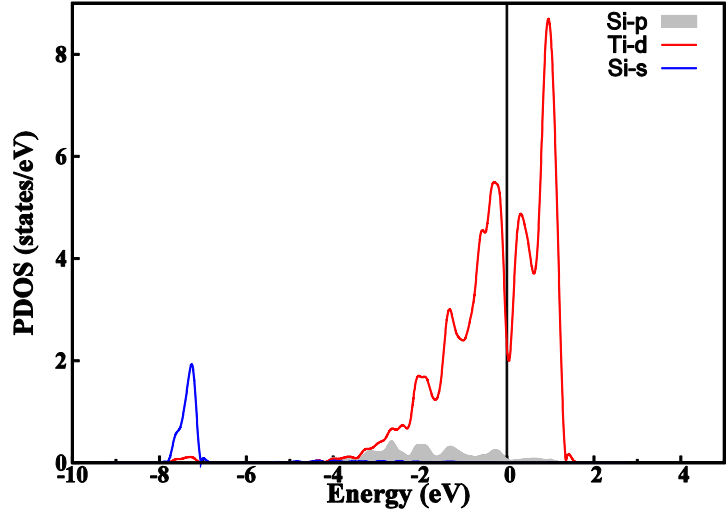

**Figure 4** Plot for the density of states (DOS) of '*s*' and '*p*' states Si atom and the '*d*' states of surface Ti atoms. The plot is for Si segregated Ti(0001) surface with 25 at.% Si. The Fermi level is set to zero.

To understand the Si distribution on the Ti(0001) surface, we considered a (4x4) supercell of Ti(0001) surface. We then substituted two Ti atoms with Si atoms. We calculated the $E_{seg}$ by varying the Si-Si distance and relaxing the systems. We observe that as the Si-Si distance was increased from 3.25 Å to 4.96 Å, the $E_{seg}$ changed from -0.48 eV/Si to -0.57 eV/Si. This clearly shows that Si clustering on the Ti surface is not favorable and the Si atoms prefer to stay as isolated impurities on the Ti(0001) surface. It also explains the lowering of the segregation energy gain for higher Si concentrations where Si atoms may lie in close proximity. Therefore, in this work we considered simple Si segregated models where the Si atoms are about 6 Å apart. The distribution of Si atoms on the Ti surface is very similar to the Si segregation in the bulk and grain boundaries of Fe as presented in Ref. [37]. In the case of bulk Fe, the Si-Si dimer with a bond length of 2.68 Å has no energy gain and thus is not favorable. On the contrary, for low Si concentration, Fe-Si solid solution with isolated Si atoms in bulk Fe was found to be stable which is consistent with the Fe-Si phase diagram.

*(3.2) Oxygen adsorption on Ti(0001) surfaces*
In order to study oxygen adsorption on Ti(0001) surface, we considered a (3x3) supercell of Ti(0001). We placed an oxygen atom at FCC, HCP and bridge sites. For Si segregated Ti(0001) surface, we considered the same (3x3) supercell where one surface Ti atom was substituted by Si corresponding to 11.11 at.% Si. On the Si segregated Ti(0001) surface, there are two different FCC(HCP) sites: (i) the oxygen atom is bonded to three Ti atoms as in clean surface and is named as FCC1(HCP1), and (ii) the oxygen atom is bonded to one Si atom and two Ti atoms and is named as FCC2(HCP2). The adsorption energy is calculated as:

$$E_{ads} = E_{tot}[Slab + O] - (E_{tot}[Slab] + \frac{1}{2}E_{tot}[O_2] \qquad (1)$$

where $E_{tot}[Slab+O]$, $E_{tot}[Slab]$ and $E_{tot}[O_2]$ are the total energies of the Ti slab model with the adsorbed oxygen, the Ti slab model and the oxygen molecule, respectively. The total energy of the oxygen molecule was calculated by placing the oxygen molecule in a cubic box of sides 20 Å and then optimizing the interatomic separation. According to Eq. (1), more negative value of $E_{ads}$ indicates stronger binding.

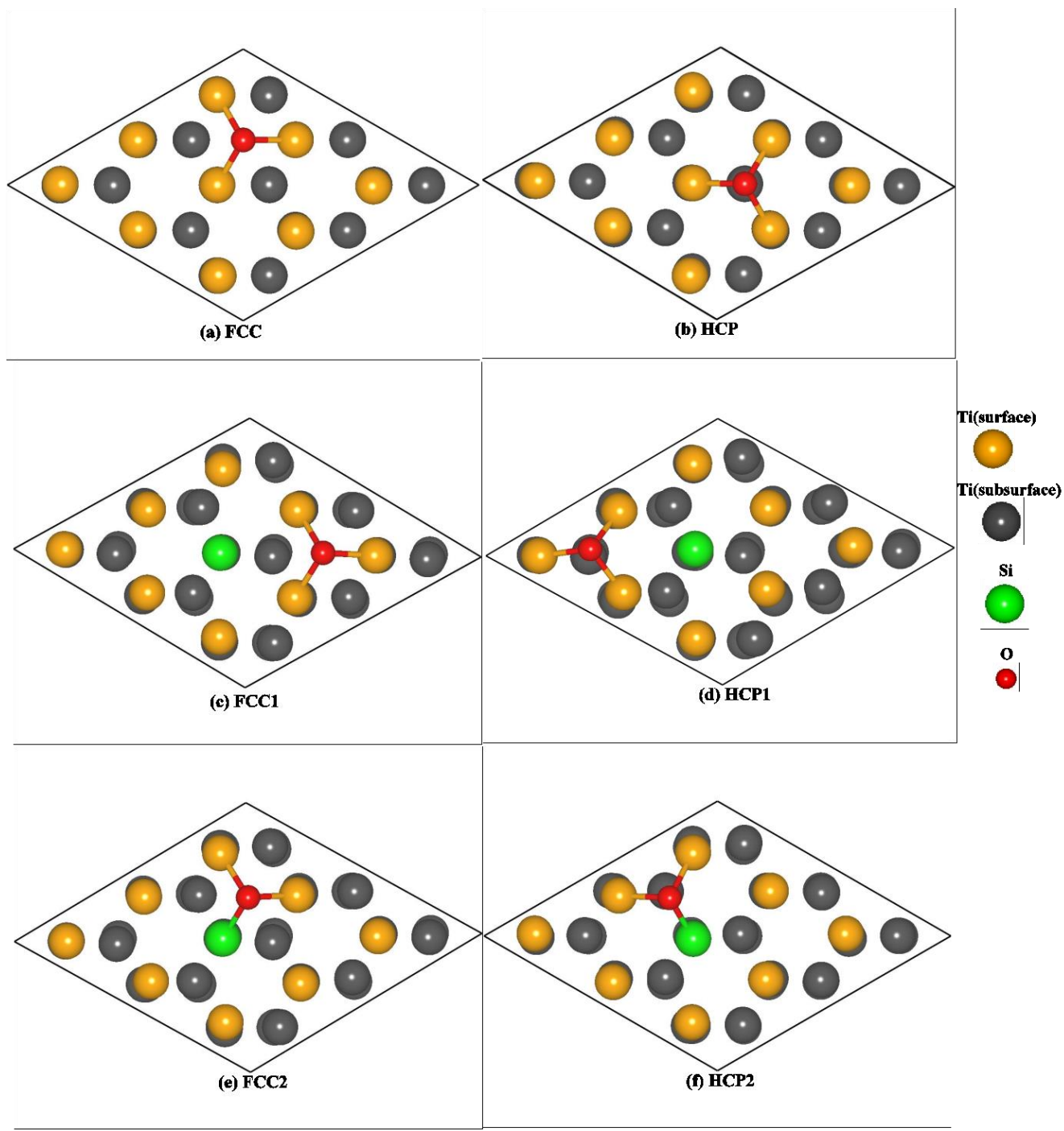

**Figure 5** Top view of different adsorption sites as well as the adsorption energy $E_{ads}$ of oxygen atom on clean and Si segregated Ti(0001) surface.

**Table 1** The oxygen adsorption energy ($E_{ads}$) and the interatomic distances between Si and O ($R_{Si-O}$) and Ti and O ($R_{Ti-O}$) for the different adsorption sites (shown in Fig. 5) are summarized below.

| Surface | Adsorption site | $E_{ads}$ (eV) | $R_{Si-O}$ (Å) | $R_{Ti-O}$ (Å) |
|---|---|---|---|---|
| Clean | FCC | -5.74 | - | 1.95 |
| Clean | HCP | -5.42 | - | 1.97 |
| Si segregated | FCC1 | -5.52 | 3.50 | 1.92, 1.98 |
| Si segregated | HCP1 | -5.29 | 3.57 | 1.92, 2.00 |
| Si segregated | FCC2 | -4.02 | 1.89 | 1.97 |
| Si segregated | HCP2 | -3.70 | 1.84 | 2.00 |

In Fig. 5, the adsorption sites for the oxygen atom on the clean and Si segregated Ti(0001) surface are shown. In Table 1 we summarize the adsorption energy for each of them. It is clear from Table 1 that the FCC site is the most favorable site on the clean Ti surface followed by the HCP site which is about 0.3 eV higher in energy. This difference in the adsorption energy between the FCC and HCP site is similar to those reported earlier [24, 32, 38]. In our calculation, we could not find any stable bridge site as the oxygen atom initially placed at the bridge site moved to the FCC site during relaxation.

In the case of Si segregated Ti(0001) surface, there are two different FCC(HCP) sites as discussed above. These are shown in Figs. 5 (c)-(f) and the $E_{ads}$ for these sites are tabulated in Table 1. The FCC1 site was found to be the most stable site followed by the HCP1 site. The difference in the adsorption energy between theFCC1 and HCP1 is about 0.3 eV which is same as that for the clean Ti(0001) surface. The FCC2 site was found to be about 1.5 eV higher in energy than the FCC1 site while the HCP2 site is the least favored site for oxygen adsorption. At the FCC1 site, even though the oxygen atom is attached to three Ti atoms, its adsorption energy is about 0.2 eV higher than the FCC site on the clean Ti(0001) surface. The Si-O distance for the FCC1 and HCP1 sites are comparable. The Si-O distance for FCC2 is 1.89 Å while the HCP2 site has the smallest Si-O distance of 1.84 Å. Interestingly for the Si segregated surface, there exists a correlation between the Si-O distance and the $E_{ads}$. The presence of Si in the neighborhood of oxygen reduces its binding with the surface atoms of Ti(0001) slab.

To understand the interaction between the oxygen atom and the Ti surfaces, we plotted the PDOS for the oxygen adsorbed clean and Si segregated Ti(0001) surfaces. These plots are shown in Fig. 6. In case of the clean Ti surface, we show the plot for the FCC site while for the Si segregated case we chose the FCC2 site. For the oxygen adsorbed clean Ti surface, we plotted the PDOS for the O-*s* and *p* states and the *d* states of the three Ti atoms bonded to the oxygen atom (see Fig. 5 (a)). The *s* states of oxygen lie very deep at around -18 eV and play little role in the bonding. The presence of oxygen generates a new peak at around -5 eV in the Ti-*d* PDOS by shifting the energy levels to the left of the Fermi level ($E_F$). A large overlap is seen between the O-*p* and Ti-*d* levels at around -5 eV. The hybridization between the O-*p* states and the Ti-*d* states gives rise to the strong bonding between the adsorbate and the adsorbent.

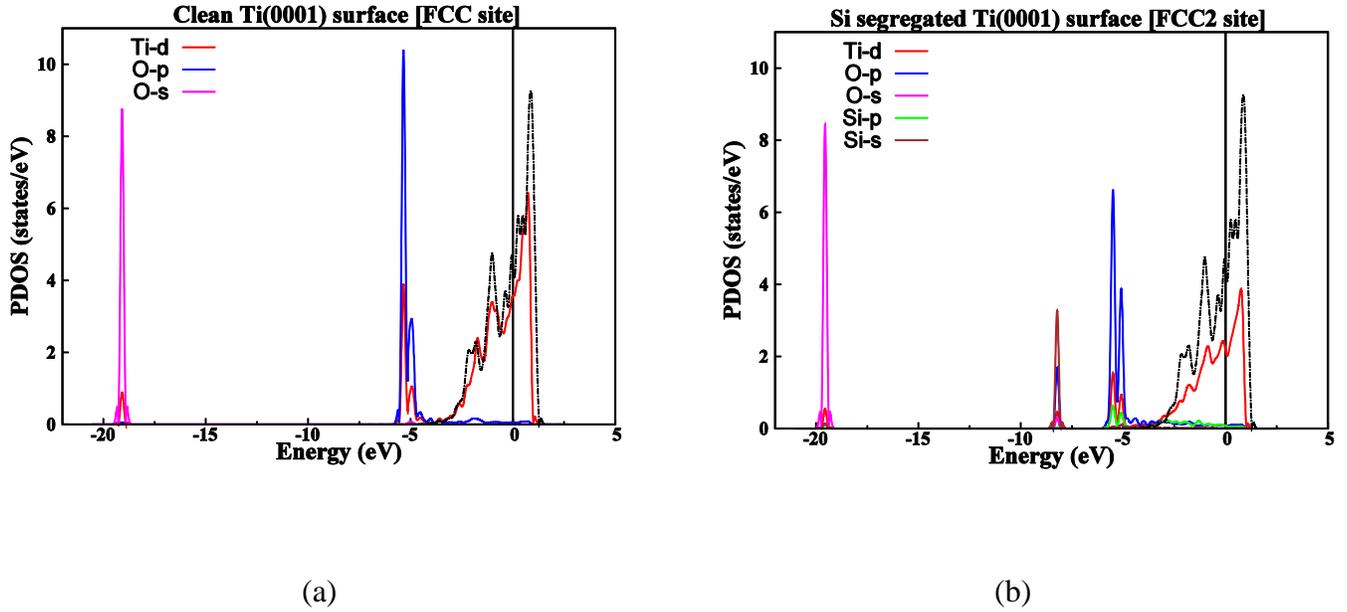

(a)                      (b)

**Figure 6** The PDOS plots for the oxygen adsorbed clean (top) and Si segregated (bottom) Ti(0001) surfaces. The Fermi level ($E_F$) is set to zero and is shown by the solid vertical line. The PDOS shown by the broken line (in both the plots) is for the Ti-$d$ states on the clean Ti(0001) surface.

In case of the Si segregated case, we plotted the PDOS for the $s$ and $p$ states of the O and Si atoms as well as the $d$ states of the two Ti atoms bonded to the oxygen atom (see Fig. 5 (e)). The O-$s$ states are very deep at around -19 eV and do not participate in the bonding, as in case of the clean Ti surface. At around -8 eV, the overlap between the Si-$s$ and O-$p$ is observed though the O-$p$ peak is small. The dominant O-$p$ peak can be seen at around -5 eV where the overlap with the Ti-$d$ and Si-$p$ states is also observed. This signifies the hybridization between the O-$p$ and Ti-$d$ states and also between $p$ states of O and Si atoms. However, the peak heights for Ti-$d$ and Si-$p$ states are very small. Notably, the Ti-$d$ peak height at around -5 eV is considerably smaller compared to the case of oxygen adsorbed clean Ti(0001) surface. The segregation of Si on Ti(0001) reduces the overlap between the Ti-$d$ and the O-$p$ states. This clearly indicates a weakening in the bonding between the adsorbate and the adsorbent.

The barrier for diffusion of the oxygen atom from the surface adsorption site to the subsurface position was calculated using the NEB method as described in section 2. The diffusion barrier is defined as the difference in the lowest energy in the initial states and the highest energy in the process when passing through the surface layer. In case of clean Ti(0001) surface the FCC site was considered as the initial site for oxygen atom while in the case of Si segregated Ti(0001) surface we considered the FCC1 and FCC2 sites. The subsurface position for the oxygen atom is the octahedral site as discussed in Refs. [24, 32, 38]. The diffusion barrier for the various initial sites is plotted in Fig. 7. The diffusion barrier for the FCC site on the clean Ti surface was calculated to be 1.28 eV. Liu et al. calculated the diffusion barrier for the FCC site to be 1.36 eV [24]. In case of the Si segregated surfaces, the barrier for the FCC1 and FCC2 sites are estimated to be 1.32 eV and 1.39 eV, respectively. Thus, Si segregation on the Ti surface increases the diffusion barrier for the oxygen atom which in turn may affect the oxygen penetration into the Ti slab. This increase in the barrier is strongly dependent on the Si-O distance. For the FCC2 site, where the Si-O distance is smaller compared to the FCC1 site, the increase in the diffusion barrier is about 0.11 eV.

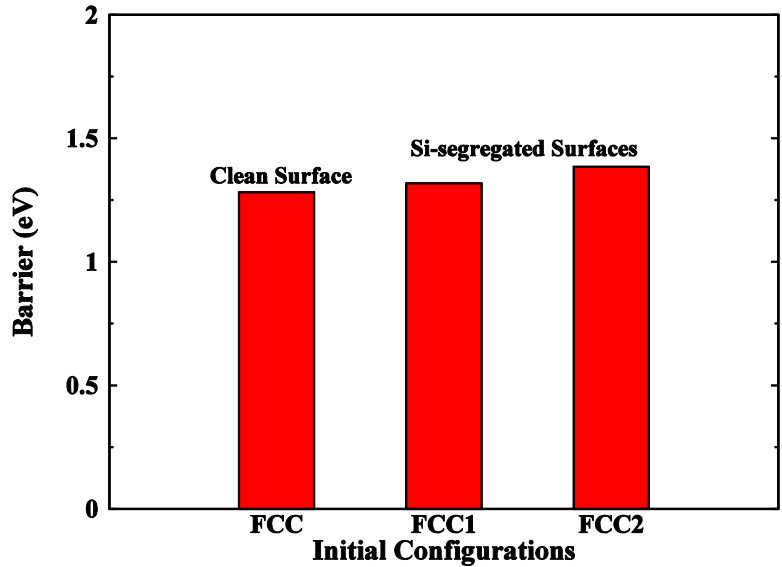

**Figure 7** The calculated energy barrier for diffusion of oxygen atom initially at various sites on the clean and Si segregated Ti(0001) surface.

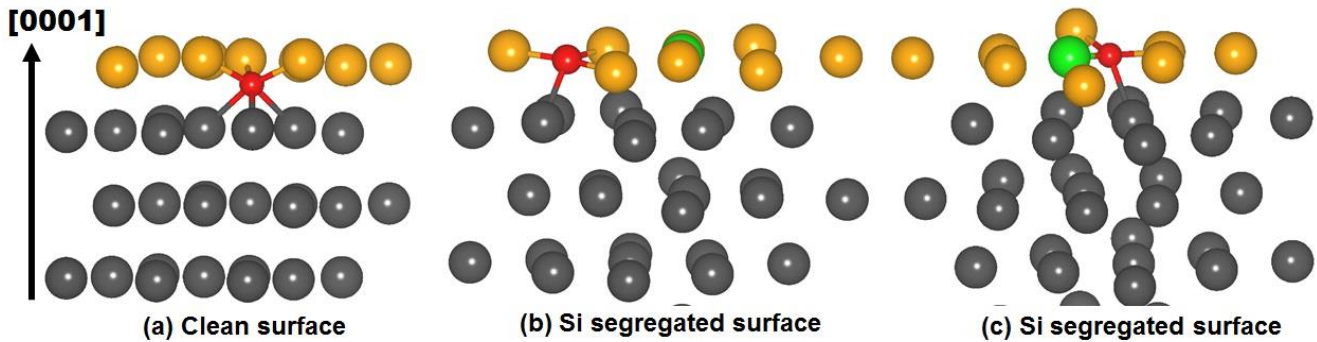

**Figure 8** The side view of the transition states (TSs) for the (a) clean surface with oxygen initially placed at the FCC site, (b) Si segregated site with oxygen initially placed at the FCC1 site, and (c) Si segregated site with oxygen initially placed at the FCC2 site. We have rotated the slabs along the [0001] plane for a better view and have shown only the relevant part. The color code is same as Fig. 5.

The difference in the diffusion barrier between the clean and Si segregated surfaces could be due to their transition states (TSs) shown in Fig. 8. A close inspection of the TS for the clean surface (see Fig. 8 (a)) revealed that the oxygen is located just below the surface and is bonded to six Ti atoms (three surface Ti atoms and three sub-surface Ti atoms). This configuration is like a distorted octahedral site where the Ti-O bond lengths vary from 2.02 Å to 2.21 Å. The TSs for the Si segregated surfaces with FCC1 and FCC2 as the initial sites are shown in Fig. 8 (b) and (c), respectively. For the FCC1 case, the oxygen atom occupies a tetrahedral site where the oxygen is bonded to four Ti atoms with the Ti-O distances in the range of 1.95 Å to 2.29 Å. For the FCC2 case, the oxygen atom occupies a tetrahedral site and is bonded to three Ti atoms (two Ti on the surface and one Ti beneath) and one Si atom. The Ti-O distances vary from 1.97 Å to 2.29 Å while the Si-O distance is 1.77 Å. Firstly, the increase in the barrier for the Si segregated Ti(0001) surfaces, compared to the clean surface, can be attributed to the observed TS where the oxygen occupy a relatively less stable tetrahedral site compared to the octahedral site in Ti [41]. Secondly, the increased barrier for the FCC2 case could be due to the close proximity of the oxygen to the Si atom which may induce further instability. All these could lead to a more unstable TS in case of the Si segregated surface thereby increasing the diffusion barrier for oxygen atoms. For Ti-Si alloys with low Si content (≤ 1 wt.%), Chaze et al. observed an inhibition of the dissolution of oxygen in Ti due to the presence of Si [16]. Additionally, they also noted that the activation energies for the oxidation of Ti-Si alloys are proportional to the Si content [17]. Hence, it is clear from previous experiments and our theoretical calculations that Si segregation hinders the dissolution of oxygen in Ti and Ti alloys.

*(3.3) Oxygen molecule adsorption on Ti(0001) surface*
In order to understand the interaction between the oxygen molecules and Ti surfaces, we adsorb an oxygen molecule on the clean and Si segregated Ti(0001) surface. We used a (4x4) supercell of Ti(0001) for adsorbing the oxygen molecule. For the Si segregated case, we considered a Ti(0001) surface with 18.75 at.% Si concentration. Initially we placed an oxygen molecule at a height of 2 Å above the surface and relaxed the structure. For the clean as well as Si segregated Ti surface, the oxygen molecule dissociates during relaxation as shown in Fig. 9. Thus, dissociation of oxygen molecule on the Ti(0001) surface is a barrier-less reaction as also observed by Liu *et al* [24]. The dissociative adsorption of the oxygen molecule on the Ti(0001) surface resembles to that on Pd(111) [39] or Al(111) [40] surfaces. For the clean Ti(0001) surface, the oxygen atoms occupy the FCC sites as evident from Fig. 9 (a). Our calculated $E_{ads}$ for the relaxed configuration yields -11.47 eV for both the oxygen atoms which is almost equal to twice that of the adsorption of a single oxygen atom at the FCC site as tabulated in Table 1. For the Si segregated case, we considered two adsorption sites for the oxygen molecule. In one case we placed it close to the Ti atom as shown in Fig. 9 (b) while in another case we placed it near the Si atom as in Fig. 9 (c). Although the oxygen molecule dissociates in both the cases, the sites occupied by the oxygen atoms are different. In case of Fig. 9 (b), which we also named as Si segregated Surface (I), one of the oxygen atoms occupy the FCC1 site while the other occupies the HCP1 site. The calculated $E_{ads}$ is -11.05 eV for both the oxygen atoms which is about 0.2 eV smaller than the sum of the adsorption energy for a single oxygen atom at the FCC1 and HCP1 sites (see Table 1). In case of Fig. 9 (c), which we labeled as Si segregated Surface (II), the oxygen atoms occupy the FCC1 and FCC2 sites. The $E_{ads}$ for this case is -9.60 eV for both the oxygen atoms which is about 0.06 eV smaller than the sum of the adsorption energy for a single oxygen atom at the FCC1 and FCC2 sites (see Table 1). However, the adsorption energies for the relaxed configurations in Figs. 9 (b) and (c) are larger compared to that of the clean surface shown in Fig. 9 (a). Overall the oxygen binding strength is weakened for the Si segregated case as compared to the clean surface.

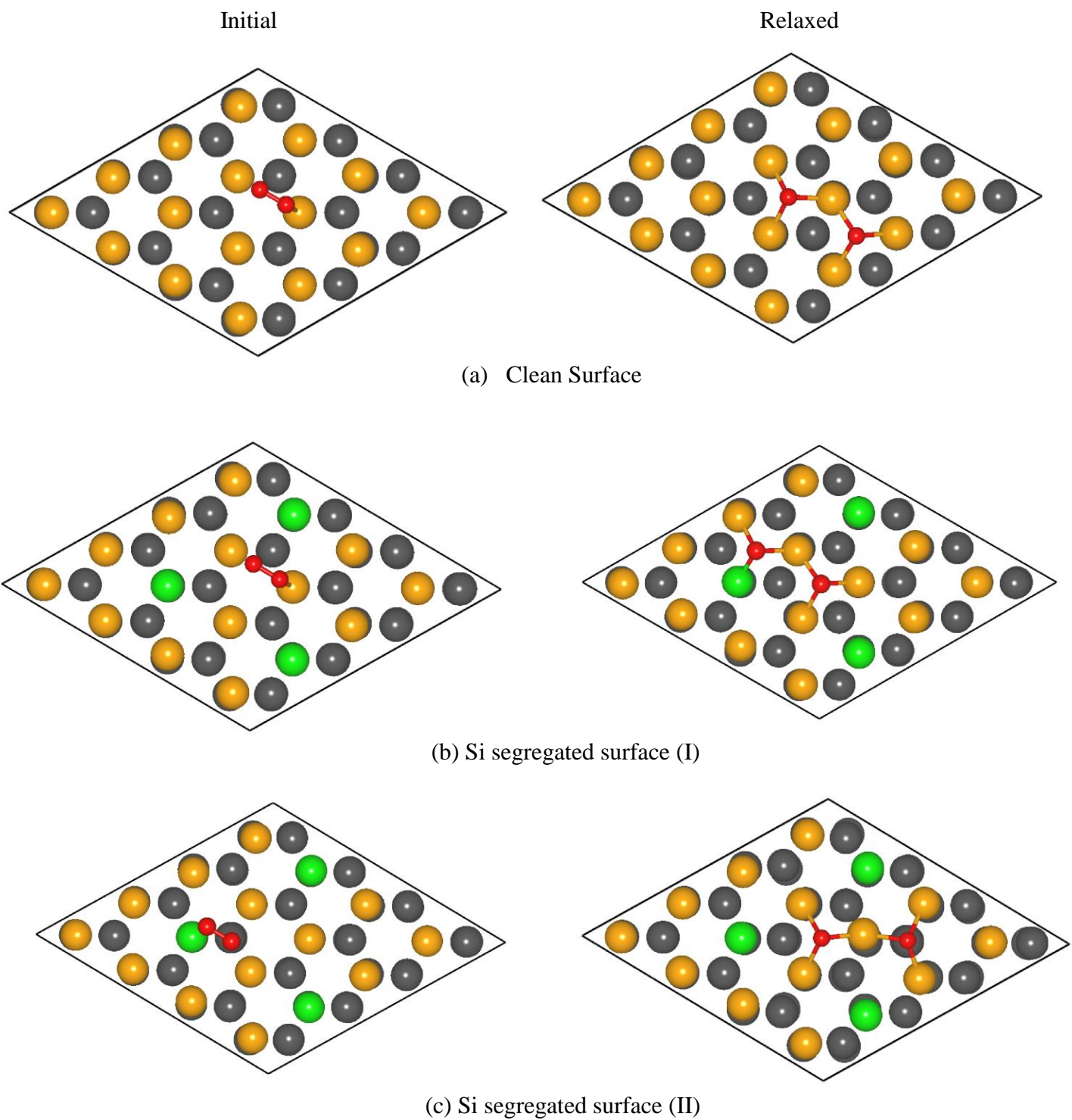

**Figure 9** The dissociation of O$_2$ molecule on the clean (a) and Si segregated (b), (c) Ti(0001) surface is shown. Only the top view of the initial and relaxed configurations are shown. The color code is same as Fig. 5.

**(4) Conclusion**

We presented our *ab initio* results of the oxygen adsorption on the clean and Si segregated Ti(0001) surface. Si atom was found to have a favorable segregation on the surface layer of Ti(0001). We calculated the Si solubility in the Ti(0001) surface to be about 75 at.%. In presence of Si, the adsorption of oxygen on Ti(0001) surface is weakened compared to the clean Ti(0001) surface. The PDOS plots explain the higher stability of those adsorption sites where the oxygen atoms form only Ti-O bonds compared to those where both Ti-O and Si-O types are formed. Dissociation of the oxygen molecule on the clean as well as Si segregated Ti(0001) surfaces was found to be barrier-less. The adsorption of oxygen atoms is weaker in case of the Si segregated surface compared to the clean surface. From our calculations, we can conclude that Ti-Si surface alloys can suppress the dissolution of oxygen in Ti by (i) reducing the binding between the oxygen and the Ti(0001) surface, and (ii) increasing the diffusion barrier for oxygen.

**Acknowledgements**

This work was supported by Council for Science, Technology and Innovation (CSTI), Cross ministerial Strategic Innovation Promotion Program (SIP), "Process Innovation for Super Heat-resistant Metals (PRISM)" (Funding agency: JST). We are extremely thankful to the high performance computing facilities in National Institute for Material Science (NIMS), Tsukuba and Institute for Materials Research (IMR), Tohoku University, Sendai for allowing us to use their resources.

**References**

1. Klinger MM, Rahemtulla F, Prince CW, Lucas LC and Lemons JE (1998), Crit. Rev. Oral Biol. Med. 9:449

2. Wallace TA, Titanium '95, Science and Technology, 1943-1950, Institute of Materials, London (1996)

3. Lütjering G and Williams JC, Titanium, 2nd ed., Engineering Materials and Processes, Springer, Berlin (2007)

4. Fu YQ, Loh NL, Batchelor AW, Liu D, Zhu XD, He J and Xu K (1998) Improvement in fretting wear and fatigue resistance of Ti-6Al-4V by application of several surface treatments and coatings. Surf. Coat. Technol. 106:193-197

5. Sonoda T and Kato M (1997) Effects of discharge voltage on Ti-O film formation on Ti-6Al-4V alloy by reactive DC sputtering. Thin Solid Films 303:196-199

6. Kim TS, Park YG, and Wey MY (2003) Characterization of Ti-6Al-4V alloy modified by plasma carburizing process. Mat. Sci. Engg. A 361:275-280

7. Kim YZ, Murakami T, Narushima T, Iguchi Y and Ouchi C (2006) Surface Hardening Treatment for C.P. Titanium and Titanium Alloys in Use of Ar-5%CO Gas. ISIJ International 46:1329-1338

8. Stringer J (1960) The oxidation of titanium in oxygen at high temperatures. Acta Metall. 8:758-766

9. Kofstad P, Anderson PB and Krudtaa OJ (1961) Oxidation of titanium in the temperature range 800-1200°C. J. Less-Common Metals 3:89-97

10. Shih HD and Jona F (1977) Low-Energy Electron Diffraction and Auger Electron Spectroscopy Study of the Oxidation of Ti {0001} at Room Temperature. Appl. Phys. 12:311-315

11. Cichy H and Fromm E (1991) Oxidation Kinetics of Metal Films at 300K Studied by the Piezoelectric Quartz Crystal Microbalance Technique. Thin Solid Films 195: 147-158

12. Lu G, Bernasek SL and Schwartz J (2000) Oxidation of a polycrystalline titanium surface by oxygen and water. Surf. Sci. 458:80-90

13. David D, Beranger G and Garcia EA (1983) A Study of the Diffusion of Oxygen in $\alpha$-Titanium Oxidized in the Temperature Range 460ºC – 700ºC. J. Electrochem. Soc. 130:1423-1426

14. Kitashima T, Yamabe-Mitarai Y, Iwasaki and Kuroda S (2016) Effects of Alloying Elements on the Tensile and Oxidation Properties of ALPHA and Near-ALPHA Ti Alloys. Proc. of the 13th World Conference on Titanium, TMS 479-483

15. Kitashima T and Kawamura T (2016) Prediction of oxidation behavior of near-α titanium alloys. Scripta Materialia 124:56–58

16. Chaze AM and Coddet C (1987) Influence of alloying elements on the dissolution of oxygen in the metallic phase during the oxidation of titanium alloys. J Mat. Sci. 22:1206-1214

17. Chaze AM and Coddet C (1987) Influence of Silicon on the Oxidation of Titanium between 550 and 700ºC. Oxidation of Metals 27:1-20

18. Vojtěch D, Bártová B and Kubatík T (2003) High temperature oxidation of titanium–silicon alloys. Mat. Sci. Engg. A 361:50-57

19. Dai J, Zhu J, Chen C and Weng F (2016) High temperature oxidation behavior and research status of modifications on improving high temperature oxidation resistance of titanium alloys and titanium aluminides: A review J. Alloys Compounds 685:784-798

20. Schneider J and Ciacchi LC (2010) First principles and classical modeling of the oxidized titanium (0001) surface. Surf. Sci. 604:1105-1115

21. Ohler B, Prada S, Pacchioni G and Langel W (2012) DFT Simulations of Titanium Oxide Films on Titanium Metal. J. Phys. Chem. C 117:358-367

22. Wu HH and Trinkle DR (2011) Direct Diffusion through Interpenetrating Networks: Oxygen in Titanium. Phys. Rev. Lett. 107:045504

23. Wu HH and Trinkle DR (2013) Solute effect on oxygen diffusion in $\alpha$-titanium. J. App. Phys. 113:223504

24. Liu J, Fan X, Sun C and Zhu W (2016) Oxidation of the titanium(0001) surface: diffusion processes of oxygen from DFT. RSC Adv. 6:71311-71318

25. Kresse G, Furthmüller J (1996) Efficiency of ab-initio total energy calculations for metals and semiconductors using a planewave basis set. Comput Mater Sci 6:15-50

26. Kresse G, Furthmüller J (1996) Efficient iterative schemes for ab initio total energy calculations using a plane-wave basis set. Phys Rev B 54:11169-11189

27. Blöchl PE (1994) Projector augmented-wave method. Phys Rev B 50:17953-17979

28. Kresse G, Joubert D (1999) From ultrasoft pseudopotentials to the projector augmented wave method. Phys Rev B 59:1758-1775

29. Perdew JP, Burke K, Ernzerhof M (1996) Generalized gradient approximation made simple. Phys Rev Lett 77:3865-3868

30. Monkhorst HJ, Pack JD (1976) Special points for Brillouin-zone integrations. Phys Rev B 13:5188-5192

31. Huda MN and Kleinman L (2005) Density functional calculations of the influence of hydrogen adsorption on the surface relaxation of Ti(0001). Phys. Rev. B 71:241406

32. Liu SY, Wang FH, Zhou YS and Shang JX (2007) Ab initio study of oxygen adsorption on the Ti(0001) surface. J. Phys.: Condens. Matter 19:226004

33. Shih H, Jona F, Jepsen D and Marcus P (1976) The structure of the clean Ti (0001) surface. J. Phys. C: Solid State Phys. 9:1405

34. Jónsson H, Mills G and Jacobsen KW, Nudged Elastic Band Method for Finding Minimum Energy Paths of Transitions, in Classical and Quantum Dynamics in Condensed Phase Simulations, Ed. B. J. Berne, G. Ciccotti and D. F. Coker, World Scientific (1998)

35. Løvvik OM (2005) Surface segregation in palladium based alloys from density-functional calculations. Surf. Sc. 583:100-106

36. Ponomareva AV, Isaev EI, Skorodumova NV, Vekilov YK and Abrikosov IA (2007) Surface segregation energy in bcc Fe-rich Fe-Cr alloys. Phys. Rev. B 75:245406

37. Bhattacharya SK, Kohyama M, Tanaka S and Shiihara Y (2014) Si segregation at Fe grain boundaries analyzed by *ab initio* local energy and local stress. J. Phys. Condens. Matter 26:355005

38. Li L, Meng F, Tian H, Hu X, Zheng W and Sun CQ (2015) Oxygenation mediating the valence density-of-states and work function of Ti(0001) skin. Phys. Chem. Chem. Phys. 17:9867

39. Todorova M, Reuter K and Scheffler M (2004) Oxygen overlayers on Pd(111) studied by density functional theory. J. Phys. Chem. B 108:14477

40. Kiejna A and Lundqvist BI (2001) First-principles study of surface and subsurface O structures at Al (111). Phys. Rev. B 63:085405.

41. Conrad H (1981) Effect of interstitial solutes on the strength and ductility of titanium. Prog. Mater. Sci. 26:123.